\documentclass[10pt,conference]{IEEEtran} 
\IEEEoverridecommandlockouts
\usepackage{cite}
\usepackage{amsmath,amssymb,amsfonts}
\usepackage{algorithmic}
\usepackage{graphicx}
\usepackage{textcomp}
\usepackage{xcolor}
\usepackage{enumitem}
\usepackage{multirow}
\usepackage{booktabs}
\newcommand{\best}[1]{\textbf{#1}}
\newcommand{\sol}{CAT-RAG}
\usepackage{booktabs}
\usepackage{makecell}

\usepackage{listings}

\usepackage{pifont}
\usepackage{framed}
\usepackage{soul}
\usepackage{booktabs}  
\usepackage{csquotes}
\usepackage{multirow}
\usepackage{array}
\newcolumntype{L}[1]{>{\raggedright\let\newline\\\arraybackslash\hspace{0pt}}m{#1}}
\newcolumntype{C}[1]{>{\centering\let\newline\\\arraybackslash\hspace{0pt}}m{#1}}
\newcolumntype{R}[1]{>{\raggedleft\let\newline\\\arraybackslash\hspace{0pt}}m{#1}}
\usepackage{url}
\MakeOuterQuote{"}
\usepackage[caption=false,font=footnotesize,labelformat=simple]{subfig} 

\usepackage[bookmarks=false]{hyperref}

\usepackage{comment}

\definecolor{mygreen}{rgb}{0,0.6,0}
\definecolor{mygray}{rgb}{0.5,0.5,0.5}
\definecolor{mymauve}{rgb}{0.58,0,0.82}

\makeatletter
\newcommand\notsotiny{\@setfontsize\notsotiny\@vipt\@viipt}
\makeatother

\usepackage[bookmarks=false]{hyperref}

\begin{document}

\title{
Improving Secure Code Generation with Category-Aware Agentic Retrieval
}
\author{\IEEEauthorblockN{Mahdieh Abazar}
\IEEEauthorblockA{\textit{University of Calgary} \\
Mahdieh.abazar@ucalgary.ca}
\and
\IEEEauthorblockN{Lorenzo De Carli}
\IEEEauthorblockA{\textit{University of Calgary} \\
\textit{
lorenzo.decarli@ucalgary.ca}}
\and
\IEEEauthorblockN{ Benjamin Tan}
\IEEEauthorblockA{\textit{University of Calgary} \\
\textit{
benjamin.tan1@ucalgary.ca
}}
%
}
\maketitle

\begin{abstract}

Despite rapid progress in LLM-based code generation, security vulnerabilities remain widespread, and retrieval-augmented code generation (RACG) can further amplify risk when retrieved knowledge is verbose or vulnerable to poisoning. Existing secure RACG approaches inject security knowledge at a fine-grained vulnerability level, causing prompt overhead and diluted model attention when a query implicates multiple related weaknesses. We propose \sol{}, an agentic security category-aware retrieval-augmented code generation framework. \sol{} performs adaptive category selection via an LLM as a categorizer agent and operates either as a standalone security layer or as a drop-in add-on to RACG pipelines, requiring no modification to the underlying model. We replace CWE-level retrieval with compact, language-specific security categories. These categories provide a higher-level abstraction over individual CWE weaknesses, organizing them by the security concern they share, such as "Input Validation and Injection Attacks," rather than merely grouping related entries. Each category consolidates best practices and secure code examples into a single, prompt-efficient representation.
We perform a comprehensive evaluation across CyberSecEval, CodeGuard+, and HumanEval. On CyberSecEval, \sol{} nearly doubles the secure code generation improvement over the state of the art (23.57\% vs. 12.11\%) while reducing token overhead by 41.71\%.

\end{abstract}


\section{Introduction}
\label{sec:introduction}

Empirical studies report that a substantial fraction of large language model (LLM)-generated code contains security weaknesses~\cite{surveyCWEllm, yan2025guiding, asleep, 
10.1007/s10664-024-10590-1}, even when functionally correct~\cite{10.1007/s10664-024-10590-1}. Retrieval-augmented code generation (RACG) enriches prompts with external knowledge retrieved at inference time, such as by retrieving functional examples that improve generated-code correctness and adding these to the input prompt~\cite{CODERAG-BENCH, survey-codegeneration}. However, retrieval is not uniformly beneficial. 
The quality, integration, and amount of context all matter, and irrelevant context can "distract" the model into poorer output~\cite{lost}. RACG knowledge bases (KBs) can also be of variable security quality, and in the worst case, be deliberately compromised through poisoning~\cite{CodeGuarder, lin2025exploringsecuritythreatsknowledge}.

Hence, existing retrieval-based approaches to improve the security of generated code, such as SecGuide\cite{11186069}, CodeGuarder~\cite{CodeGuarder}, SOSecure~\cite{mukherjee2025sosecure}, SecCoder~\cite{zhang2024seccoder}, and RESCUE~\cite{rescue}, inject security knowledge, drawn from Common Weakness Enumeration (CWE) entries, Stack Overflow discussions, or curated corpora to discourage insecure patterns. 
These systems organize security knowledge at the granularity of individual CWEs or vulnerability instances.

However, a single user query can implicate multiple related vulnerabilities~\cite{asleep}, e.g., a memory-allocation task simultaneously touching buffer overflows, integer overflows, use-after-free, and null-pointer dereferences. 
This creates two problems for per-CWE retrieval. 
First, retrieving an entry per weakness produces long, verbose prompts that dilute model attention and can harm generation quality~\cite{lost, REPOFORMER}. 
Second, developers tend to reason at the level of programming tasks and API usage~\cite{nadi_jumping_2016}, rather than abstract security concepts such as CWEs: requests are expressed at a task-oriented level spanning broad domains (e.g., reading a file, loading it into memory, and processing it involves input handling, memory management, and file operations). This does not necessarily correspond to a single identifier such as CWE-119 (buffer overflow).

The gap between CWE-level retrieval granularity and how developers naturally reason about tasks, together with the need for concise, focused context, motivates our core hypothesis: vulnerability types cluster naturally by programming domain, so grouping information about related CWEs into compact, language-specific categories for RACG should both reduce prompt overhead and better align with developer intent. Thus, we propose \sol{}, which organizes security knowledge into categories, each combining best practices and secure code examples that cover multiple related CWEs, and uses a lightweight LLM as an agent to adaptively select relevant categories for each query. \sol{} requires no model changes and works as a drop-in security layer atop standalone generation or existing RACG pipelines. Overall, this paper offers the following contributions:

\begin{itemize}
    \item We propose language-specific, category-based security knowledge for RACG that groups related CWEs by shared security concern rather than by individual identifier into compact programming-domain entries for C, C++, Python, and Java, reducing prompt overhead compared to prior work while preserving actionable guidance.
    \item We design an adaptive retrieval mechanism that selects a variable number of categories per query based on its complexity, rather than applying a fixed retrieval budget. 
    \item We evaluate \sol{} on CyberSecEval \cite{bhatt2023purplellamacybersecevalsecure} across standalone, functional-RAG, summarization, poisoning, and retrieval-strategy settings, and on CodeGuard+ \cite{codeguard+} and HumanEval \cite{humaneval} under standalone settings. We focus on open-weight, low-cost code models as they represent an affordable, private approach to code generation~\cite{MAHADEHASAN2026112815} yet they often generate insecure code~\cite{explicit}. Our results demonstrate improved security rates over CodeGuarder and improved resilience to both targeted and intent-agnostic KB poisoning.

\end{itemize}

\section{Background and Related Work}
\label{sec:background}

\subsubsection*{Code Generation and Retrieval}
Code-oriented LLMs~\cite{hui2024qwen2, zhu2024deepseek, roziere2023code} have advanced rapidly due to improvements in model architecture, training techniques, and access to large-scale code corpora. While many models demonstrate strong performance on a wide range of programming tasks~\cite{roziere2023code}, empirical results suggest a persistent security gap: LLM-generated code frequently contains vulnerabilities, even when functionally adequate~\cite{10.1007/s10664-024-10590-1, CodeSecEval}. RACG~\cite{CODERAG-BENCH, tao2025retrieval, REDCODER} addresses functional limitations by grounding model outputs in up-to-date, domain-specific external knowledge.

\subsubsection*{Security Weaknesses of LLM-Generated Code}
Intrinsic weaknesses in LLM-generated code arise from training data and model architecture,  manifesting even without external retrieval. 
Recent studies highlight that issues such as null pointer dereferences, buffer overflows, and array bounds violations occur frequently~\cite{surveyCWEllm, yan2025guiding}. 
 Conventional defenses rely on scanning code with static analysis tools that can flag insecure patterns (e.g., Semgrep\cite{semgrep_tool}, Weggli~\cite{ weggli_tool}, or CodeQL~\cite{codeql}). However, such tools suffer from false positives\cite{11500264}, and do not address the problem at the root (i.e., during generation).

\subsubsection*{Model-Centric Hardening}
These approaches work without external retrieval. SVEN~\cite{sven} uses prefix-based control to steer outputs, and in\-struc\-tion-tuning methods~\cite{SafeCoder, li2024finetuningsecure, li2026exploratory} improve secure generation via curated corpora. More recent work~\cite{hasan2025teaching} applies localized preference optimization to target security-relevant code. These methods often require fine-tuning or specialized, not widely-accessible infrastructure~\cite{cosec,CoSEFA}.

\subsubsection*{Security-Oriented RACG}

Security-oriented RACG is also effective for improving secure code generation. SecGuide~\cite{11186069} retrieves task-specific guidelines from MITRE CWE and CodeQL, comparing RAG setup against recursive self-critique; it finds the two comparable, with their combination further reducing insecure code. SOSecure~\cite{mukherjee2025sosecure} instead operates post-generation, retrieving Stack Overflow answers and comments that flag vulnerabilities in similar code and feeding them back as revision context. RESCUE~\cite{rescue} targets the retrieval pipeline itself, building a distilled hierarchical security KB and using multi-faceted retrieval to capture task-relevant security semantics rather than noisy raw documents.
However, at the high-level stage, usually only a natural-language query is available, so RESCUE must first produce a code draft solely to infer relevant CWEs, incurring the overhead of an extra generation step whose CWE inference is only as reliable as that initial, potentially insecure draft.
SecCoder~\cite{zhang2024seccoder} uses retrieved safe demonstrations for in-context learning, helping models apply vulnerability-fix patterns more robustly and generalize to unseen or attacked settings.
CodeGuarder~\cite{CodeGuarder}, the leading state-of-the-art approach, retrieves vul\-ner\-a\-bil\-i\-ty-aware knowledge alongside functional examples: it decomposes user queries into sub-tasks via a multi-stage, fine-grained mechanism, maps each segment to specific CWE entries and curated secure patterns, and re-ranks and filters this knowledge by the host LLM's susceptibility to the identified vulnerability types before injecting it into the prompt. However, this per-CWE granularity produces long, verbose prompts when a query spans multiple interrelated concerns, diluting model attention and potentially harming code quality. It is also not aligned with how developers reason (i.e., in terms of broad domains such as ``memory handling'' rather than individual CWE identifiers).
For our side-by-side comparisons with prior related work, we focus on CodeGuarder as it reports the best security enhancement among the prior work so far.

RACG grounds LLM outputs in external, domain-specific knowledge to improve functional correctness~\cite{CODERAG-BENCH, tao2025retrieval}, but the security quality of that KB is not guaranteed; integrating external data can itself open new attack vectors. Poisoned or vulnerable KBs systematically bias models toward insecure patterns~\cite{lin2025exploringsecuritythreatsknowledge, asleep}, and in the worst case, adversaries might inject semantically similar vulnerable examples that rank highly at retrieval time~\cite{lin2025exploringsecuritythreatsknowledge}. This is especially concerning because functional KBs are often built from public repositories like GitHub~\cite{lin2025exploringsecuritythreatsknowledge} or internal codebases integrated without rigorous security validation. We therefore also evaluate our approach in a poisoning setting, following prior work\cite{CodeGuarder}, to characterize the worst-case scenario in which security-oriented RACG operates alongside a poisoned functionality-oriented KB.

\setlength{\FrameSep}{2pt}

\begin{figure*}[t]
  \centerline{
  \includegraphics[width=0.9\linewidth]{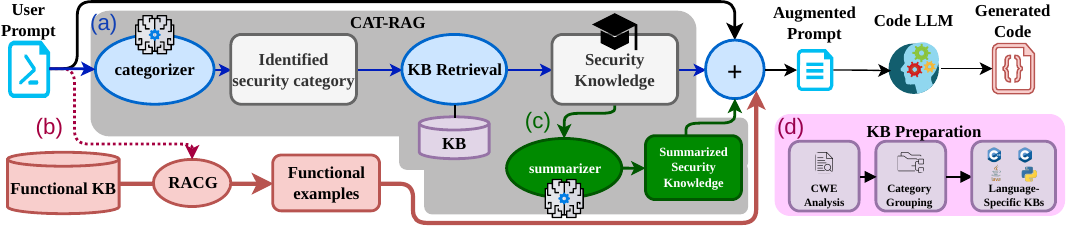}}
  \caption{Overview of the CAT-RAG framework. (a) Core category-aware security path (blue): the categorizer selects categories, retrieves matching knowledge, and augments the prompt. (b) Functional-RACG path (red), retrieving functional code examples. (c) Summarization variant (green), compacting retrieved security knowledge. (d) Offline KB preparation (pink inset): CWE analysis, category grouping, and language-specific KB construction.}
  \label{fig:cat-rag}
\end{figure*}

\section{Outline of Proposed Approach}\label{sec:method} 

We propose \textit{\sol}, an agentic category-aware retrieval-augmented code generation framework that provides security knowledge in compact and domain-based programming context categories to harden LLM code generation.
Rather than aggregating CWEs directly, these categories are organized around broad programming and security domains such as memory operations, file handling, and cryptography. Given a query, the system identifies the most relevant domains and retrieves the associated best practices and secure code examples as unified categories.
As discussed in section \ref{sec:introduction}, we are motivated by the observation that vulnerabilities cluster by programming domain.
This design lets \sol{} serve as an agentic security layer, cutting token usage via compact categories and avoiding context budget exhaustion.
\sol{} guides LLMs by identifying different possible security threats and giving hints for mitigating or avoiding such threats. Fig.~\ref{fig:cat-rag} is an overview of \sol{}.

\subsection{Category-Aware Retrieval}
Given a user query, this phase identifies the relevant security categories and constructs the augmented prompt. We use an LLM to read the query, identify the target programming language, and select the most relevant categories from the corresponding language-specific KB (discussed later). Path (a) in Fig.~\ref{fig:cat-rag} shows the flow of this phase.

\subsubsection{%
Categorization}
We propose using an LLM as a categorizer agent, since LLMs are well-suited to identifying the semantic intent of user queries. Given a query, the categorizer is prompted with the full list of category names across all supported programming languages and asked to return the detected language and a ranked list of relevant category IDs.
Importantly, we propose \textit{adaptive} category selection, which means the number of identified categories can vary. Rather than returning a fixed top-k set of categories, the categorizer can "adapt" its selection (from $\sim$10 categories per language) in response to the query. A query
might be mapped to a single security category,
while a complex query involving file I/O, memory allocation, and string manipulation might correspond, for example, to three or more. 
This adaptive behavior, which we call "$dynamic~k$", helps reduce context usage; simpler queries can receive less injected knowledge, conserving context budget, while complex queries can draw on more categories spanning the multiple domains they touch.

\sol{} also includes a summarization step (path (c) in Fig.~\ref{fig:cat-rag}). It is optional because it adds an extra inference call: when enabled, an LLM-based summarizer shortens the retrieved best practices and code examples into a more compact, potentially more effective form, at the cost of that additional call. We explore this variation to see the effect of making retrieved knowledge more concise.

\begin{table}[t]
\caption{Security categories per programming language.}
\label{tab:categories}
\centering
\scriptsize
\setlength{\tabcolsep}{4pt}
\renewcommand{\arraystretch}{1.1}
\begin{tabular}{@{}l p{6.6cm}@{}}
\toprule
\textbf{Lang} & \textbf{Categories} \\
\midrule
C & Memory \& Buffer Ops; Input Validation \& Injection; System Commands \& Processes; File \& Path Ops; Authentication \& Access Control; Encryption \& RNG; Resource Limits \& DoS; Sensitive Info Protection; File Permissions \& Privileges; HTTP \& URL Handling \\
\midrule
C++ & Memory Safety \& RAII; Input Validation \& Injection; Exception Safety \& Error Handling; Type Safety \& Casting; Concurrency \& Thread Safety; STL \& Container Safety; Cryptography \& RNG; Move Semantics \& Ownership; File \& Stream Ops; Template \& Compile-Time Safety \\
\midrule
Python & Input Validation \& Injection; OS Cmd Execution; File Ops \& Path Traversal; Deserialization \& Data Parsing; Authentication \& Session Mgmt; Cryptography \& RNG; Error Handling \& Info Disclosure; HTTP Security \& Web Apps; SSRF; Resource Mgmt \& DoS \\
\midrule
Java & Input Validation \& Injection; OS Cmd Execution; File Ops \& Path Traversal; Deserialization Security; Authentication \& Session Mgmt; Cryptography \& RNG; Error Handling \& Info Disclosure; HTTP Security \& Web Apps; SSRF; Resource Mgmt \& DoS \\
\bottomrule
\end{tabular}
\end{table}

\begin{figure}[t]
\centering
\setlength{\fboxsep}{4pt}
\fbox{%
  \begin{minipage}{0.92\columnwidth}
    \footnotesize
    \textbf{File \& Path Operations - Best Practices}\\[3pt]
    $\bullet$ Canonicalize and validate all file paths\\[1pt]
    $\bullet$ Restrict access to allowed base directories\\[1pt]
    $\bullet$ Check file type and size before processing\\[1pt]
    $\bullet$ Use \texttt{O\_NOFOLLOW} to prevent symlink attacks\\[1pt]
    $\bullet$ Never trust user-provided filenames
  \end{minipage}%
}\\[4pt]
\begin{minipage}{0.96\columnwidth}
\begin{scriptsize}
\begin{verbatim}
#include <TRUNCATED FOR SPACE IN THE PAPER...>
/* Path traversal prevention (CWE-22/434) */
int safe_open(const char *base,
              const char *user_path) {
  char full[PATH_MAX],
       cb[PATH_MAX], cf[PATH_MAX];
  if (!realpath(base, cb)) return -1;
  snprintf(full, sizeof(full),
           "%s/%s", cb, user_path);
  if (!realpath(full, cf)) return -1;
  /* CWE-22 */
  if (strncmp(cb, cf, strlen(cb)) != 0)
    return -1;
  /* CWE-59 */
  int fd = open(cf, O_RDONLY|O_NOFOLLOW);
  if (fd < 0) return -1;
  struct stat st;
  /* CWE-434 */
  if (fstat(fd,&st)||!S_ISREG(st.st_mode)
                   ||st.st_size>10485760){
    close(fd); return -1;
  }
  return fd;
}
\end{verbatim}
\end{scriptsize}
\end{minipage}
\caption{Security best practices for the "File \& Path Operations"
  category (C language KB entry) (top), and secure file-open helper (C)
  with guards for CWE-22 (path traversal), CWE-59 (symlink following),
  and CWE-434 (unrestricted upload) (bottom). The C example is truncated for space reasons here, but the whole KB is available in the paper's artifacts.}
\label{fig:file-path-combined}
\end{figure}

\subsubsection{Prompt Construction}
Once the relevant categories are retrieved, \sol{} constructs the security-augmented prompt by appending each of the selected category's best practices and secure code examples to the original user query.

\subsection{Constructing Language-specific KBs}

Our approach involves categorizing/compacting language-specific KB entries to address multiple CWEs in order to reduce context overhead (path (d) in Fig.~\ref{fig:cat-rag} ). 
Not all security knowledge is equally useful for code generation; language-agnostic guidance risks confusing the model by introducing advice inappropriate for the target language, a problem exacerbated when retrieved context is long or noisy~\cite{lost}.

In this work, we construct separate KBs for C, C++, Python, and Java.
We constructed the weakness categories through a manual, expert-driven analysis of vulnerability patterns. This process started by gathering the MITRE Top-25 most dangerous software weaknesses \cite{mitre_cwe_top25_2025} and findings from an empirical study on vulnerabilities in LLM-generated code~\cite{10.1007/s10664-024-10590-1}. 
It reports that null pointers, buffer overflows, invalid pointer errors, and array bounds violations are among the most frequent weaknesses produced by LLMs. 
Based on this information, the expert defined a set of common programming contexts, such as file handling, memory operations, input validation, and cryptographic usage, where such weaknesses often occur. 
The relevant CWEs were then assigned to the contexts in which they are most likely to appear. Some CWEs were linked to multiple contexts because they can occur in different situations. This mapping resulted in a flexible and practical categorization scheme.
Table~\ref{tab:categories} shows the category names based on this process. In our implementation, some domains are shared across languages (e.g., input validation and cryptography), and others are specific to a language's paradigm (e.g., deserialization for Python and Java). This helps the retrieved security knowledge include advice "native" to the target language, avoiding the confusion that can arise from language-agnostic retrieval.

Finally, for each category, we created short best-practice guidelines and simple code examples with the help of an LLM.
The guidelines were written as concise, actionable bullet points, and the code examples were kept minimal to clearly show how to prevent the related CWEs without adding unnecessary complexity. 

\subsubsection*{Category Schema}
Each category is a structured entry with the following fields; Fig.~\ref{fig:file-path-combined} shows an example.
\begin{itemize}
  \item\textbf{Primary Name and ID} captures the category name and a numeric identifier. The LLM-based categorizer uses the name to extract related categories, while the ID serves as a retrieval index for matching user queries to categories.
  \item\textbf{Best Practices} provide a concise, actionable set of security guidelines derived from CWEs, written as informative statements. These guidelines are short to cover essential security problems while keeping length low.
  \item\textbf{Secure Code Example} provides a compact, language-specific example showing a correct implementation of secure practices. We annotate each with CWE references to show the LLM which parts relate to the CWE.
\end{itemize}

\noindent As we will see in Section~\ref{sec:results}, these categories provide an effective KB which brings significant improvements over the state of the art for function-level queries. We emphasize that the process of building this KB was independent of the benchmarks and relied only on the process explained earlier in this section. However, we also stress that these categories only represent an initial exploration of this space, and we do not cover all CWEs and complex cases. Our main contribution here is the formulation and evaluation of the concept of category-based knowledge and agentic retrieval.

\section{Evaluation Methodology}\label{sec:study}
To evaluate the performance of \sol, we answer the following research questions: 
\begin{enumerate}[label=\textbf{RQ\arabic*},leftmargin=*]
\item \textbf{Effectiveness:} How effective is \sol{} at improving code security in standalone/functional RACG settings?
\item \textbf{Resilience:} How resilient is \sol{} under knowledge base poisoning attacks? (This captures RACG in scenarios where the functional KB is less-than-ideal for security). 
\item \textbf{Design Sensitivity:} How do retrieval depth, categorizer choice, and knowledge base architecture affect \sol's performance?
\item \textbf{Retrieval Strategy:} How does LLM-based category retrieval compare against naive retrieval baselines over the same knowledge base?
\end{enumerate}

\subsection{Investigated Scenarios}

\sol{} is an independent security-hardening layer that can slot into different code-generation workflows depending on deployment context and resources available. 
We evaluate it (i) standalone, with a coding LLM (path (a) in Fig. 1); (ii) on top of an existing functional RACG pipeline that first retrieves a code example from a functional KB (paths (a)+(b)); and (iii) with an added LLM summarizer over the retrieved best practices and code examples (paths (a)+(c)). Together, these address effectiveness across use cases (RQ1), sensitivity to design choices (RQ3), and the advantage of LLM-based category retrieval over naive baselines (RQ4). Our baseline throughout is an LLM generating without any retrieval.

For RQ2, we adopt the threat models of prior work~\cite{CodeGuarder, lin2025exploringsecuritythreatsknowledge}, assuming the attacker knows the code-generation framework and targets the \textit{functional} RACG KB. In Scenario I (exposed intents), the attacker knows the user's query and injects vulnerable examples most similar to it, so the poisoned code ranks highly at retrieval. In Scenario II (intent-agnostic), the attacker lacks the query and instead poisons broadly: it clusters the KB, takes the top 10\% most representative entries (centroids likely retrieved across many queries), and injects their closest vulnerable counterparts. 
Both represent a pessimistic case in which a functionality-oriented KB contains deliberately security-misleading content, thus also generalizing to the more optimistic case where the KB has unintentional security-misleading content (such as outdated, buggy code).

\subsection{Benchmarks}
We adopt CyberSecEval~\cite{bhatt2023purplellamacybersecevalsecure} as our main evaluation benchmark, a comprehensive benchmark developed by Meta as part of the Purple Llama project.
It evaluates LLMs in two domains: their propensity to generate insecure code and their compliance with cyberattack requests. For checking insecure code, CyberSecEval uses the Insecure Code Detector (ICD), which has 189 rules based on Semgrep \cite{semgrep_tool} and Weggli \cite{weggli_tool} to find 50 types of insecure coding practices defined by CWE across eight programming languages, including C, C++, Python, and Java. Prior work also uses this benchmark (i.e., CodeGuarder~\cite{CodeGuarder}, the state-of-the-art approach we compare against). 
We focus on the four languages supported by \sol's KBs: C (227 instances), C++ (259 instances), Java (229 instances), and Python (351 instances), for a total of 1,066 instances (spanning 33 distinct CWEs). 

We further probe whether \sol{}'s gain generalizes. In particular, recent work argues that security and functional correctness should be evaluated jointly~\cite{rethinking}. To do so, we evaluate \sol{} on CodeGuard+~\cite{codeguard+}, a secure code generation benchmark for Python and C/C++ used in recent RAG-based work such as RESCUE~\cite{rescue}. 
Unlike CyberSecEval~\cite{bhatt2023purplellamacybersecevalsecure}, CodeGuard+~\cite{codeguard+} pairs each instance with functional test cases and evaluates security using CodeQL~\cite{codeql}, allowing us to assess security and functional correctness in a single setting. 
Using the benchmark's official repository, we obtain 103 instances spanning 42 distinct CWE labels.

Finally, to isolate \sol{}'s impact on functional correctness, we use HumanEval~\cite{humaneval}, a standard benchmark for Python generation, comprising 164 problems focusing on algorithmic tasks, math, and edge cases, and evaluates correctness via unit tests. Unlike CyberSecEval and CodeGuard+, HumanEval carries no security dimension, making it a clean probe for any functionality regression. 
Readers can find details about the benchmarks in their respective papers~\cite{bhatt2023purplellamacybersecevalsecure, codeguard+, humaneval}.

\subsection{Evaluation Metrics}
We adopt metrics that are well-established in the secure code generation literature~\cite{bhatt2023purplellamacybersecevalsecure, codeguard+, CodeGuarder}: we evaluate security using Security Rate ($SR$), code similarity using $CodeBLEU$, functional correctness using $pass@k$, and functional and security jointly using $secure-pass@k$.

\subsubsection{Security Rate (SR)~\cite{sven}}
We use SR as our primary metric, corresponding to the insecure coding practice pass rate defined in both CyberSecEval and CodeGuard+. SR is the fraction of generated code samples in which no insecure coding practices are detected by the static analysis tool (Semgrep + Weggli for CyberSecEval and CodeQL for CodeGuard+).%

\subsubsection{CodeBLEU~\cite{ren2020codebleumethodautomaticevaluation}}
As CyberSecEval does not have executable test cases for functional correctness,
we follow prior state-of-the-art work~\cite{CodeGuarder} and use CodeBLEU, which extends BLEU with AST and data-flow matching components, to compare generated code against reference code. We report this as a secondary metric to measure how security improvements affect functional similarity. Due to this limitation, we use Pass@k and Secure-Pass@k for further evaluation.

\subsubsection{Pass@k~\cite{humaneval}}
In code generation benchmarks such as HumanEval and CodeGuard+, the pass@k metric characterizes functional correctness; we use it to be consistent with prior work. It is computed by sampling $k$ independent completions per problem and checking whether at least one of them passes all unit tests. Formally, 
\(
  \text{pass@}k = \mathbb{E}_{x \in X}\left[1 - \frac{\binom{n_x - c_x}{k}}{\binom{n_x}{k}}\right],
\)
where $n_x$ is the number of generated samples for problem $x$, $c_x$ is the number of those samples that pass the unit tests, and the expectation is taken over all problems in the dataset $X$~\cite{humaneval}. For pass@1, the single generated solution must satisfy every test case; for pass@5, the model passes if any of five samples is fully correct.

\begin{table*}[t]
\centering
\caption{Standalone CAT-RAG (Mistral Categorizer, $k=1$--$3$) (left), CAT-RAG + Functional RACG (Mistral Categorizer, $k=1$--$3$, ReposVul) (middle), CAT-RAG + Summarization (GPT-5-mini Categorizer/Summarizer, $k=1$--$5$) (right). Best SR change $\Delta$\% per Language in Bold per experiment.
\label{tab:standalone}\label{tab:pipeline2}\label{tab:summarization}
}
\resizebox{\linewidth}{!}{%
\begin{tabular}{@{}llrrrrrr|rrrrrr|rrrrrr@{}}
\toprule
 & & \multicolumn{6}{c}{\textbf{STANDALONE CAT-RAG}} & \multicolumn{6}{c}{\textbf{CAT-RAG + FUNCTIONAL RACG}} & \multicolumn{6}{c}{\textbf{CAT-RAG + SUMMARIZATION}} \\
 & & \multicolumn{3}{c}{\textbf{SR (\%)}} & \multicolumn{3}{c}{\textbf{CodeBleu (\%)}} & \multicolumn{3}{c}{\textbf{SR (\%)}} & \multicolumn{3}{c}{\textbf{CodeBleu (\%)}} & \multicolumn{3}{c}{\textbf{SR (\%)}} & \multicolumn{3}{c}{\textbf{CodeBleu (\%)}} \\ \cmidrule(lr){3-5} \cmidrule(lr){6-8} \cmidrule(lr){9-11} \cmidrule(lr){12-14} \cmidrule(lr){15-17} \cmidrule(lr){18-20}
\textbf{Model} & \textbf{Lang} & Base    & CAT     & $\Delta$\%       & Base       & CAT       & $\Delta$\%         & Base    & CAT     & $\Delta$\%       & Base       & CAT       & $\Delta$\%         & Base    & CAT     & $\Delta$\%       & Base      & CAT       & $\Delta$\%         \\ \midrule
\multirow{4}{*}{\textbf{Qwen-Coder}}     & C             & 55.51   & 66.96   & 20.63            & 21.57      & 22.6      & 4.78               & 55.51   & 70.04   & 26.18            & 21.57      & 22.53     & 4.45               & 55.51   & 69.16   & 24.59            & 21.57     & 22.77     & 5.56               \\
               & C++           & 74.52   & 86.49   & 16.06            & 23.84      & 24.19     & 1.47               & 74.52   & 86.1    & 15.54            & 23.84      & 24.28     & 1.85               & 74.52   & 87.26   & 17.1             & 23.84     & 24.47     & 2.64               \\
               & Java          & 44.1    & 53.71   & 21.79            & 27.6       & 27.08     & $-$1.88              & 44.1    & 52.4    & 18.82            & 27.6       & 27.06     & $-$1.96     & 44.1    & 50.22   & 13.88            & 27.6      & 26.89     & $-$2.57              \\
               & Py            & 68.66   & 76.64   & \textbf{11.62}   & 19.88      & 21.68     & 9.05               & 68.66   & 78.35   & \textbf{14.11}   & 19.88      & 21.67     & 9.00                  & 68.66   & 77.21   & \textbf{12.45}   & 19.88     & 21.12     & 6.24               \\ \midrule
\multirow{4}{*}{\textbf{DS-Coder}}       & C             & 52.86   & 68.72   & \textbf{30.0}      & 21.89      & 23.15     & 5.76               & 52.86   & 67.4    & 27.51            & 21.89      & 23.43     & 7.04      & 52.86   & 74.01   & 40.01            & 21.89     & 24.39     & 11.42              \\
               & C++           & 71.43   & 84.94   & 18.91            & 23.6       & 24.45     & 3.6                & 71.43   & 85.33   & 19.46            & 23.6       & 25.04     & 6.1       & 71.43   & 89.19   & 24.86            & 23.6      & 25.58     & 8.39               \\
               & Java          & 42.79   & 62.88   & \textbf{46.95}   & 27.01      & 26.24     & $-$2.85            & 42.79   & 63.32   & \textbf{47.98}   & 27.01      & 26.28     & $-$2.7               & 42.79   & 64.63   & \textbf{51.04}   & 27.01     & 27.29     & 1.04               \\
               & Py            & 72.36   & 75.78   & 4.73             & 20.94      & 23.7      & 13.18     & 72.36   & 75.78   & 4.73             & 20.94      & 23.02     & 9.93      & 72.36   & 77.78   & 7.49             & 20.94     & 23.08     & 10.22              \\ \midrule
\multirow{4}{*}{\textbf{CodeLlama}}      & C             & 60.35   & 77.53   & 28.47            & 19.39      & 21.51     & 10.93     & 60.35   & 79.74   & \textbf{32.13}   & 19.39      & 20.73     & 6.91               & 60.35   & 84.58   & \textbf{40.15}   & 19.39     & 22.38     & 15.42     \\
               & C++           & 70.66   & 88.03   & \textbf{24.58}   & 20.7       & 22.77     & 10        & 70.66   & 87.26   & \textbf{23.49}   & 20.7       & 21.58     & 4.25               & 70.66   & 90.73   & \textbf{28.4}    & 20.7      & 22.66     & 9.47      \\
               & Java          & 57.21   & 77.73   & 35.87            & 22.36      & 23.29     & 4.16      & 57.21   & 82.1    & 43.51            & 22.36      & 20.11     & $-$10.06             & 57.21   & 72.93   & 27.48            & 22.36     & 23.9      & 6.89      \\
               & Py            & 75.78   & 82.62   & 9.03             & 18.62      & 20.55     & 10.37              & 75.78   & 82.91   & 9.41             & 18.62      & 19.41     & 4.24               & 75.78   & 78.63   & 3.76             & 18.62     & 21.63     & 16.17     \\ 
               \bottomrule
\end{tabular}%
}
\end{table*}

\subsubsection{Secure-Pass@k~\cite{codeguard+}}
For benchmarking with CodeGuard+, we report secure-pass@k, which measures the probability that at least one of the $k$ generated samples is both functionally correct and secure. Unlike pass@k, this metric requires a generated sample to satisfy \textit{both} unit tests and static security checks. We use secure-pass@k to see if \sol{} improves security without losing ability to write working code.

\subsection{Studied LLMs}
We use three code-oriented LLMs for code generation: Qwen2.5-Coder-7B~\cite{hui2024qwen2}, DeepSeek-Coder-V2-16B~\cite{zhu2024deepseek}, and CodeLlama-13B \cite{roziere2023code}. 
For categorization, we primarily use Mistral-7B~\cite{jiang2023mistral7b} (deployed locally via Ollama), and 
GPT-5-mini \cite{singh2025openai} as an alternative categorizer. 
For summarization, we also explore GPT-5-mini and Mistral. 
For brevity, we refer to these models as Qwen-Coder (Q.C.), DS-Coder (D.C.), CodeLlama (C.L.), and Mistral throughout the remainder of this paper.
Our choice of LLMs aims at ensuring representation of a variety of open-weight model sizes; these also represent models that developers might choose to run locally.

\section{Experimental Results}\label{sec:results}

This section presents the experimental evaluation of \sol{} organized by our four research questions. 
The improvement percentages ($\Delta$\%) are computed relative to the baseline for each model and language. 
Note, in poisoning scenarios, the baseline is poisoned RACG, and improvement percentages are computed relative to this baseline.
For experiments related to CodeGuard+, we report Secure-Pass@1, Pass@1 and SR. In the other experiments for evaluating the functional correctness related to HumanEval, we report Pass@1 and Pass@5. The highest improvement per language across models is shown in bold. For space, CAT is used instead of CAT-RAG in tables.

\subsection{Experimental Setup}

We experiment using a server with NVIDIA A100 GPU and use the Ollama framework for LLMs, except for GPT-5-mini (accessed via the OpenAI API). 
For Mistral as categorizer, we set the temperature to 0.1, and for consistent and reproducible generated codes across all LLMs, we generally set the temperature parameter to 0.
For evaluating pass@5 with HumanEval, we set the temperature to 0.8 to generate different samples. 
We set maximum number of generated tokens (MAX\_NEW\_TOKEN) to 4096 for CyberSecEval and 2048 for CodeGuard+ and HumanEval, while keeping all other parameters at their defaults. 
Generally, we used a common prompt format across all models per benchmark for fairness (e.g., for HumanEval, we used a code completion prompt, while in CyberSecEval, the prompt was for code generation).

Implementation details and source code for CAT-RAG are available in our repository.
For functional RACG and poisoning scenarios, we build on the implementation provided by prior work~\cite{CodeGuarder}, using their codebase to run both the RACG pipeline and the CodeBLEU evaluation of LLM-generated responses on CyberSecEval. The functional RACG setup is adopted as-is from this prior work.%

\subsection{RQ1: Effectiveness}

We evaluate the standalone \sol{} pipeline and measure the security of the generated code. 
We also evaluate it on top of typical RACG, where it acts as a security-oriented retrieval layer. The ``fixed'' code in ReposVul~\cite{reposvul} is used as the functionality-focused KB. 
We then extend the standalone \sol{} with a summarization step to investigate the effect of more precise information on guiding LLMs. 
Finally, we assess whether security-oriented retrieval adversely affects functional correctness, focusing on standalone \sol. 

\subsubsection{Standalone \sol{}}
Table~\ref{tab:standalone} (left) presents results using Mistral as the categorizer with $dynamic~k\!=\!1-3$. \sol{} improves SR across all 12 model-language combinations without exception, with an overall improvement of 22.39\% on average. 
We observe that larger models tend to benefit more: DS-Coder (16B) achieves the largest improvement (+25.15\%), followed by CodeLlama (13B, +24.49\%), while Qwen-Coder, the smallest model (7B), shows the lowest (+17.53\%).
Across languages, gains are largest for Java (+34.87\%), whose baseline security is also lowest, leaving more room for improvement followed by C (+26.37\%) and C++ (+19.85\%), with Python the most modest (+8.46\%). CodeBLEU remains stable or improves in 10 of 12 configurations.

\subsubsection{\sol{} + Functional RACG}
Table~\ref{tab:pipeline2} (middle) presents results using Mistral as the categorizer when \sol{} operates on top of a functional RACG.
The average SR improvement is 23.57\% slightly exceeding the standalone setting.  The cross-language ranking mirrors the standalone results, with Java again showing the largest gains (+36.77\%), the highest Java SR across all \sol{} variants. This indicates that \sol{} provides additive security value on top of functional retrieval. We hypothesize that the functional examples help the LLM understand implementation patterns, while \sol{} independently injects security guidance that steers the generated code toward secure alternatives. 
The CodeBLEU drop seen in some cases does not necessarily indicate worse code, since it may simply diverge in style from the reference. This motivated us to further evaluate \sol{} on functional benchmarks such as CodeGuard+ and HumanEval.

\subsubsection{\sol{} + Summarization}

Adding an LLM-based summarization step (GPT-5-mini as both categorizer and summarizer, in two separate calls) yields an average SR improvement of +24.27\% (Table~\ref{tab:standalone} (right)), modestly above standalone CAT-RAG. The effect is language-dependent: C and C++ improve over standalone (+26.37\% to +34.92\% and +19.85\% to +23.45\%), whereas gains shrink for Java (+34.87\% to +30.8\%) and Python (+8.46\% to +7.9\%), indicating that summarization does not uniformly help across languages. Consistent with the standalone setting, larger models tend to benefit more. CodeBLEU also improves over the standalone setting.

\subsubsection{Impact on Functional Correctness}
\label{sec:functional}
Table~\ref{tab:codeguard} reports secure-pass@1, pass@1, and SR for the standalone pipeline with Mistral on CodeGuard+. \sol{} improves secure-pass@1 for DS-Coder and CodeLlama, with CodeLlama gaining most across all three metrics. DS-Coder improves on secure-pass@1 and pass@1 while SR is unchanged. Qwen-Coder shows the opposite pattern: pass@1 and secure-pass@1 both fall even as SR rises slightly, indicating a minor trade-off for this model, though its pass@1 was already near the ceiling, leaving little room to improve.

We also evaluate functional correctness on HumanEval~\cite{humaneval}. Based on Table~\ref{tab:functional}, \sol{} improves both pass@1 and pass@5 across all three models, indicating that CAT-RAG preserves and can even enhance functional correctness.

\begin{table}[t]
\caption{CodeGuard+ Results.
Standalone CAT-RAG (Mistral Categorizer, $k\!=\!1$--$3$), Greedy Decoding. Bold Indicates Improvement over Baseline. 
}
\label{tab:codeguard}
\centering
\resizebox{\columnwidth}{!}{%
\begin{tabular}{l rrr rrr rrr}
\toprule
& \multicolumn{3}{c}{\textbf{Secure-Pass@1 (\%)}} & \multicolumn{3}{c}{\textbf{Pass@1 (\%)}} & \multicolumn{3}{c}{\textbf{SR (\%)}} \\
\cmidrule(lr){2-4} \cmidrule(lr){5-7} \cmidrule(lr){8-10}
\textbf{Model} & Base & CAT & $\Delta$\% & Base & CAT & $\Delta$\% & Base & CAT & $\Delta$\% \\
\midrule
Q.C. & 62.13 & 59.22 & $-$4.68 & 98.05 & 94.17 & $-$3.96 & 62.57 & 64.36 & \best{$+$2.86} \\
D.C. & 34.95 & 44.66 & \best{$+$27.78} & 65.04 & 66.99 & \best{$+$3.00} & 64.07 & 64.07 & 0.00 \\
C.L. & 34.95 & 45.63 & \best{$+$30.56} & 63.10 & 73.78 & \best{$+$16.93} & 55.33 & 70.95 & \best{$+$28.23} \\
\bottomrule
\end{tabular}%
}
\end{table}

\begin{table}[t]
\caption{Impact on Functional Correctness on HumanEval (Mistral Categorizer). Bold Indicates Higher Value}
\label{tab:functional}
\centering
\scriptsize
\setlength{\tabcolsep}{5pt}
\begin{tabular}{l rrr rrr}
\toprule
& \multicolumn{3}{c}{\textbf{pass@1}} & \multicolumn{3}{c}{\textbf{pass@5}} \\
\cmidrule(lr){2-4} \cmidrule(lr){5-7}
\textbf{Model} & Base & CAT & $\Delta$\% & Base & CAT & $\Delta$\% \\
\midrule
Qwen-Coder & 74.39 & 81.71 & \best{$+$9.84} & 88.41 & 89.63 & \best{$+$1.38} \\
DS-Coder   & 72.56 & 73.78 & \best{$+$1.68} & 82.32 & 84.15 & \best{$+$2.22} \\
CodeLlama  & 41.46 & 45.12 & \best{$+$8.83} & 59.15 & 61.59 & \best{$+$4.13} \\
\bottomrule
\end{tabular}
\vspace{-1em}
\end{table}

\begin{table}[t]
\caption{%
  \sol{} (Mistral categorizer, $k{=}1$--$3$) vs.\ CodeGuarder (CG)~\cite{CodeGuarder}
  on CyberSecEval, same ReposVul KB and baseline.
  \textit{Top}: SR(\%) and imp. ($\Delta\%$) over
  baseline; bold marks the higher $\Delta\%$ per pair.
  \textit{Bottom}: avg.\ prompt tokens per query (input / output / total);
  \textit{saves} is \sol{}'s total-token reduction vs.\ CG.}
\label{tab:vs-codeguarder}
\centering
\scriptsize
\renewcommand{\arraystretch}{0.95}
\setlength{\tabcolsep}{4pt}
\resizebox{\columnwidth}{!}{%
\begin{tabular}{@{}ll r rr rr@{}}
  \toprule
  & & &
    \multicolumn{2}{c}{\textbf{\sol{}}} &
    \multicolumn{2}{c}{\textbf{CG}} \\
  \cmidrule(lr){4-5}\cmidrule(lr){6-7}
  \textbf{Model} & \textbf{Lang} & \textbf{Base} &
    \textbf{SR} & $\boldsymbol{\Delta\%}$ &
    \textbf{SR} & $\boldsymbol{\Delta\%}$ \\
  \midrule
  \multirow{4}{*}{\textbf{Qwen-Coder}}
    & C    & 55.51 & 70.04 & \best{$+$26.18} & 66.52 & $+$19.83 \\
    & C++  & 74.52 & 86.10 & \best{$+$15.54} & 82.63 & $+$10.88 \\
    & Java & 44.10 & 52.40 & $+$18.82 & 65.50 & \best{$+$48.53} \\
    & Py   & 68.66 & 78.35 & $+$14.11 & 79.20 & \best{$+$15.35} \\
  \midrule
  \multirow{4}{*}{\textbf{DS-Coder}}
    & C    & 52.86 & 67.40 & \best{$+$27.51} & 61.23 & $+$15.83 \\
    & C++  & 71.43 & 85.33 & \best{$+$19.46} & 77.99 & $+$9.18 \\
    & Java & 42.79 & 63.32 & \best{$+$47.98} & 54.15 & $+$26.55 \\
    & Py   & 72.36 & 75.78 & \best{$+$4.73} & 75.78 & \best{$+$4.73} \\
  \midrule
  \multirow{4}{*}{\textbf{CodeLlama}}
    & C    & 60.35 & 79.74 & \best{$+$32.13} & 61.23 & $+$1.46 \\
    & C++  & 70.66 & 87.26 & \best{$+$23.49} & 71.04 & $+$0.54 \\
    & Java & 57.21 & 82.10 & \best{$+$43.51} & 53.28 & $-$6.87 \\
    & Py   & 75.78 & 82.91 & \best{$+$9.41} & 75.21 & $-$0.75 \\
  \midrule
  \multicolumn{3}{@{}l}{\textit{Avg.\ $\Delta\%$ (12 pairs)}}
    & & \best{$+$23.57} & & $+$12.11 \\
  \bottomrule
\end{tabular}%
}

\smallskip

\resizebox{\columnwidth}{!}{%
\begin{tabular}{@{}l rrr rrr r@{}}
  \toprule
  & \multicolumn{3}{c}{\textbf{\sol{}}}
  & \multicolumn{3}{c}{\textbf{CodeGuarder}} & \\
  \cmidrule(lr){2-4}\cmidrule(lr){5-7}
  \textbf{Model} &
    \makecell{\textbf{In}\\Token} & \makecell{\textbf{Out}\\Token} & \makecell{\textbf{Tot.}\\Token} &
    \makecell{\textbf{In}\\Token} & \makecell{\textbf{Out}\\Token} & \makecell{\textbf{Tot.}\\Token} &
    \makecell{\textbf{CAT-RAG}\\\textbf{saves}} \\
  \midrule
  \textbf{Q.C.} & 1545.0 & 276.1 & 1821.1 & 2866.2 & 290.7 & 3156.8 & \best{$-$42.31\%} \\
  \textbf{D.C.} & 1515.8 & 369.2 & 1884.9 & 2845.8 & 316.2 & 3162.0 & \best{$-$40.39\%} \\
  \textbf{C.L.} & 1515.8 & 296.0 & 1811.7 & 2845.8 & 301.6 & 3147.4 & \best{$-$42.44\%} \\
  \midrule
  \textit{Avg.} & 1525.5 & 313.8 & \textbf{1839.2} & 2852.6 & 302.8 & \textbf{3155.4} & \best{\textbf{$-$41.71\%}} \\
  \bottomrule
\end{tabular}%
}
\vspace{-1em}
\end{table}

\subsubsection{Comparison with CodeGuarder}
\label{sec:vs-codeguarder}

We additionally compare \sol{} with CodeGuarder~\cite{CodeGuarder}, the most closely related prior work and the current state of the art in security-oriented RACG, as it reports stronger performance than SVEN \cite{sven}, SafeCoder \cite{SafeCoder}, SOSecure \cite{mukherjee2025sosecure}, CoSec \cite{cosec}, and RCI \cite{RCI}. Under matched experimental conditions, both systems augment the same code-generation LLMs with the same ReposVul functional KB and are evaluated on the CyberSecEval.
\sol{} operates with a Mistral categorizer and $dynamic~k=1-3$; CodeGuarder uses its published default parameters ($k'=2$ decomposed sub-tasks, $k=5$ CWE entries retrieved per sub-task). Table~\ref{tab:vs-codeguarder} reports SR and token cost for both.

\sol{} achieves a mean SR improvement of $+$23.57\% across all 12 model-language pairs, nearly double CodeGuarder's $+$12.11\%, surpassing it outright in 9 of 12 and tying in one. The two cases where CodeGuarder leads both involve Qwen-Coder (Java and Python); for DS-Coder and CodeLlama, \sol{} is superior in every language. CodeGuarder also degrades security below the retrieval-less baseline in two cases, CodeLlama/Java and CodeLlama/Python, whereas \sol{} stays positive across all pairs.

As discussed earlier, one of our aims was to provide security knowledge in a more efficient, compact manner.
Table~\ref{tab:vs-codeguarder} (Bottom) reports input, output, and total token counts averaged across the CyberSecEval benchmark, per model. \sol{}'s category-level abstraction is substantially more compact than CodeGuarder's multi-stage, per-CWE retrieval.
Also, \sol{} uses on average 41.71\% fewer total tokens than CodeGuarder,
with consistent savings across all three models.
The reduction is driven by the input side, \sol{} injects far fewer prompt tokens by replacing a verbose set of per-CWE entries with some compact, domain-aligned security categories, while output lengths remain broadly comparable across both systems. Consequently, \sol{} delivers greater average security improvement while operating at roughly 58\% of CodeGuarder's total token budget, leaving more context headroom for functional retrieval, longer model outputs, or deployment in context-window-constrained environments.
\begin{framed}
\textbf{RQ1 Findings:}
Without modifying the underlying LLM, CAT-RAG improves SR across all 12 model-language configurations while preserving functional correctness on HumanEval. Against CodeGuarder under matched conditions, it nearly doubles the mean SR gain (+23.57\% vs. +12.11\%) using 41.71\% fewer tokens, never dropping below the baseline.

\end{framed}

\begin{table}
  \caption{\sol{} under Poisoning Scenarios~I (exposed intents) and~II (intent-agnostic) with GPT-5-mini as categorizer and summarizer, $k=1$-$5$. Best SR $\Delta$\% per language in \textbf{bold}.}
  \label{tab:defense}
  \resizebox{\columnwidth}{!}{%
  \begin{tabular}{ll rrr rrr rrr}
    \toprule
    & & \multicolumn{3}{c}{\textbf{Qwen-Coder}} & \multicolumn{3}{c}{\textbf{DS-Coder}} & \multicolumn{3}{c}{\textbf{CodeLlama}} \\
    \cmidrule(lr){3-5} \cmidrule(lr){6-8} \cmidrule(lr){9-11}
    & \textbf{Lang} & Pois. & Def. & $\Delta$\% & Pois. & Def. & $\Delta$\% & Pois. & Def. & $\Delta$\% \\
    \midrule
    \multirow{4}{*}{\makecell{Poisoning \\Scenario I}}
    & C    & 47.58 & 55.07 & +15.74 & 46.26 & 57.71 & \textbf{+24.75} & 48.46 & 55.07 & +13.64 \\
    & C++  & 53.28 & 61.39 & +15.22 & 56.76 & 66.02 & \textbf{+16.31} & 54.83 & 59.46 & +8.44 \\
    & Java & 47.16 & 52.40 & +11.11 & 46.29 & 64.63 & \textbf{+39.62} & 52.40 & 69.87 & +33.34 \\
    & Py   & 45.01 & 54.42 & \textbf{+20.91} & 49.57 & 52.71 & +6.33 & 47.58 & 52.42 & +10.17 \\
    \midrule
    \multirow{4}{*}{\makecell{Poisoning \\Scenario II}}
    & C    & 61.67 & 67.40 & +9.29 & 58.59 & 76.21 & +30.07 & 59.91 & 82.38 & \textbf{+37.51} \\
    & C++  & 76.83 & 88.80 & +15.58 & 75.68 & 85.71 & +13.25 & 72.20 & 84.17 & \textbf{+16.58} \\
    & Java & 48.47 & 52.40 & +8.11 & 48.03 & 62.88 & +30.92 & 54.15 & 72.93 & \textbf{+34.68} \\
    & Py   & 72.36 & 77.49 & \textbf{+7.09} & 72.08 & 76.35 & +5.92 & 75.21 & 79.20 & +5.31 \\
    \bottomrule
  \end{tabular}%
  }
\end{table}

\subsection{RQ2: Poisoning Resilience}

We evaluate \sol{} under two poisoning scenarios following Lin et al.~\cite{lin2025exploringsecuritythreatsknowledge}. The \sol{} configuration uses GPT-5-mini as both categorizer and summarizer by two queries with dynamic $k\!=\!1$-$5$.

\subsubsection{Poisoning Scenario I, Exposed Intents}
The first part of Table~\ref{tab:defense} shows that \sol{} improves SR by 17.97\% on average across all 12 model–language pairs over the poisoned baseline, with every configuration improved.
Java benefits most, SR dramatically recovers +28.02\% on average across all models. 
C and C++ show consistent recovery of +18.04\% and +13.32\%, respectively, while Python recovery is 12.47\%. 

\subsubsection{Poisoning Scenario II, Intent-Agnostic}
The second part of Table~\ref{tab:defense} shows that \sol{} improves SR by 17.86\% on average over the poisoned baseline, with all 12 configurations improving. C and Java benefit most, with SR on average 25.63\% and +24.57\% respectively across all models.

\begin{framed}
\textbf{RQ2 Findings:} 
CAT-RAG recovers SR under both poisoning scenarios (+17.97\% and +17.86\%), despite Scenario I being the stronger, query-targeted attack. The explicit security signal helps override the bias from poisoned examples regardless of whether the attacker knows the target query.

\end{framed}

\subsection{RQ3: Design Sensitivity}
We investigate the design choices that affect \sol{}'s performance, examining three dimensions in this ablation study: retrieval depth (the number of categories retrieved, $k$); the choice of categorizer model and the effect of post-retrieval summarization; and KB design.

\begin{figure}[t]
  \centerline{
  \includegraphics[width=1\linewidth]{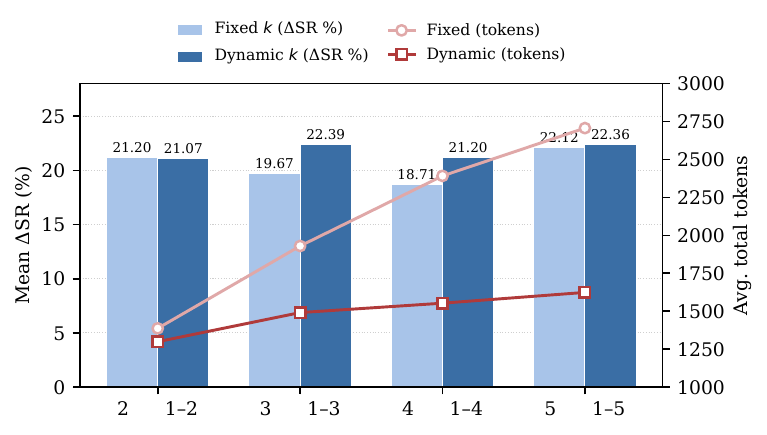}}
  \caption{Fixed vs.\ dynamic retrieval depth, averaged over all 12 model--language pairs (Mistral categorizer). Bars: mean $\Delta$SR (left axis); lines: avg.\ total tokens/query (right axis). Dynamic $k$ matches or exceeds fixed-$k$ on security while using far fewer tokens, with the gap widening at higher $k$.}
  \label{fig:k-comparison}
\end{figure}

\subsubsection{Effect of Retrieved Category Count}
A key hyperparameter in any RAG pipeline is how many "documents" to inject into a prompt. 
Here, retrieving too few categories could leave relevant security guidance uncovered, while retrieving too many could introduce noise, risking the dilution of the LLM's attention. 
To understand this trade-off, we evaluate two complementary strategies: $Fixed~k \in \{2, 3, 4, 5\}$, where a fixed number of categories is retrieved for every query regardless of its complexity, and $Dynamic~k \in \{1\text{-}2, 1\text{-}3, 1\text{-}4, 1\text{-}5\}$, where the LLM-based categorizer adaptively selects the number of categories up to the given ceiling.

Fig.~\ref{fig:k-comparison} reports SR improvement and token cost averaged across all 12 model-language configurations.
Among fixed-$k$, SR oscillates across values, making it hard to select a single $k$ that generalizes without per-deployment tuning.
On average, dynamic $k$ matches or exceeds all fixed settings.
Though the aggregate margin is narrow, dynamic $k$'s advantage is query-level adaptivity: focused queries receive only the most targeted categories, preserving prompt budget, while complex multi-domain queries get broader coverage.
Since no fixed $k$ is universally optimal,
we recommend dynamic $k{=}1$\textendash$3$ as the default: it matches the best fixed setting while keeping overhead low, consuming up to ${\approx}45\%$ fewer input tokens than fixed $k{=}5$ at the same ceiling.

\begin{table}[t]
\caption{Average $\Delta$SR~(\%) across languages and total
  token cost per query, by categorizer/summarizer configuration,
  $k{=}1$\textendash$5$. \textit{Cat.}~=~categorizer;
  \textit{Sum.}~=~summarizer.}
\label{tab:categorizer-merged}
\centering\footnotesize\setlength{\tabcolsep}{3pt}
\resizebox{\columnwidth}{!}{%
\begin{tabular}{l r r  r r  r r  r r}
\toprule
\multirow{2}{*}{\textbf{Model}} &
\multicolumn{2}{c}{\makecell{Mistral-Cat/\\Mistral-Sum}} &
\multicolumn{2}{c}{\makecell{Mistral-Cat/\\GPT-Sum}} &
\multicolumn{2}{c}{\makecell{GPT-Cat/\\Mistral-Sum}} &
\multicolumn{2}{c}{\makecell{GPT-Cat/\\GPT-Sum}} \\
\cmidrule(lr){2-3}\cmidrule(lr){4-5}\cmidrule(lr){6-7}\cmidrule(lr){8-9}
& $\Delta SR\%$ & Tot. Cost & $\Delta SR\%$ & Tot. Cost
& $\Delta SR\%$ & Tot. Cost & $\Delta SR\%$ & Tot. Cost \\
\midrule
\textbf{Q.C.}
  & +13.32 & 1046.7 & +18.47 & 1274.9
  & +14.71 & 1156.4 & +17.01 & 1354.2 \\
\textbf{D.C.}
  & +12.93 & 1114.3 & +27.27 & 1474.6
  & +22.73 & 1234.8 & +30.85 & 1574.5 \\
\textbf{C.L.}
  & +12.48 & 1068.3 & +24.95 & 1387.3
  & +17.85 & 1231.0 & +24.95 & 1568.7 \\
\bottomrule
\end{tabular}%
}
\end{table}

\subsubsection{Effect of Categorizer and Summarizer Choice}

We evaluate two categorizer/summarizer candidates: Mistral, a lightweight open-source model, and GPT-5-mini, a commercial model with per-token charges.
We ask whether the added cost of summarization is justified by security gains, and whether a fully local pipeline (Mistral-Cat/Mistral-Sum) can match a commercial alternative.
Table~\ref{tab:categorizer-merged} reports $\Delta$SR\% and Total Token Cost across four combinations at $k{=}1$\textendash$5$, denoted \textit{X-Cat/Y-Sum} for the categorization and summarization models, respectively.
Since summarizing is optional over the categorizer-only pipeline, our comparison is twofold: whether it improves SR over the baseline, and whether it adds SR gains beyond the standalone categorizer to justify added cost.

Under this lens, Mistral-based summarization (both Mistral-Cat/Mistral-Sum and GPT-Cat/Mistral-Sum) largely fails to justify the added cost: it yields the two lowest mean $\Delta$SR\% values (+12.91\% and +18.43\%), not improving on its GPT-summarizer counterpart at comparable categorizer cost.
In contrast, GPT-5-mini summarization offers a better trade-off in both configurations that use it.
GPT-Cat/GPT-Sum attains the highest mean SR improvement of any configuration (+24.27\% across the three models), at an average total cost of ${\sim}$1,499 tokens per query, comparable to the dynamic $k{=}1$\textendash$3$ categorizer-only setting (Fig.~\ref{fig:k-comparison}, ${\sim}$1,491) but better SR.

Mistral-Cat/GPT-Sum offers an even more attractive operating point: a +23.56\% mean SR improvement at only ${\sim}$1,379 tokens per query, lower than the dynamic $k{=}1$\textendash$3$ categorizer-only setting (Fig.~\ref{fig:k-comparison}) and giving the best cost–performance trade-off.
A locally hosted categorizer keeps inference cost minimal, while GPT-5-mini summarization turns the added call into a SR gain.

\begin{table}[t]
\caption{KB architecture comparison across four configurations (Mistral,
  $k=1-5$): \textit{Lang-Specific KBs} (\sol{} default), \textit{Lang-Specific KBs. Best Practices Only}, \textit{Unified KB with Cross-Lang.\ Best Practices and Code}, and \textit{Unified KB with Cross-Lang.\ Best Practices Only}. SR in~\%. Best $\Delta\%$ per language in \best{bold}.}
\label{tab:kb-arch}
\centering\footnotesize\setlength{\tabcolsep}{4pt}
\resizebox{\columnwidth}{!}{%
\begin{tabular}{l l  r r r  r r r  r r r}
\toprule
\multirow{2}{*}{\textbf{Configuration}} &
\multirow{2}{*}{\textbf{Lang}} &
\multicolumn{3}{c}{\textbf{Qwen-Coder}} &
\multicolumn{3}{c}{\textbf{DS-Coder}} &
\multicolumn{3}{c}{\textbf{CodeLlama}} \\
\cmidrule(lr){3-5}\cmidrule(lr){6-8}\cmidrule(lr){9-11}
& & Base & SR & $\Delta\%$ & Base & SR & $\Delta\%$ & Base & SR & $\Delta\%$ \\
\midrule
\multirow{4}{*}{\makecell[l]{\textit{Lang-Specific KBs} \\\textit{(Best Practices + Code)}\\\textit{(\sol{} Default)}}}
  & C    & 55.51 & 66.96 & \best{+20.63} & 52.86 & 69.16 & \best{+30.84} & 60.35 & 77.09 & \best{+27.74} \\
  & C++  & 74.52 & 87.64 & \best{+17.61} & 71.43 & 84.94 & \best{+18.91} & 70.66 & 89.96 & \best{+27.31} \\
  & Java & 44.10 & 54.15 & \best{+22.79} & 42.79 & 61.40 & \best{+43.49} & 57.21 & 76.86 & \best{+34.35} \\
  & Py   & 68.66 & 76.64 & \best{+11.62} & 72.36 & 76.35 & +5.51         & 75.78 & 81.48 & +7.52  \\
\midrule

\multirow{4}{*}{\makecell[l]{\textit{Lang-Specific KBs}\\\textit{(Best Practices Only)}}}
  & C    & 55.51 & 61.67 & +11.10 & 52.86 & 51.10 & -3.33 & 60.35 & 61.23 & +1.46 \\
  & C++  & 74.52 & 79.92 & +7.25 & 71.43 & 77.99 & +9.18 & 70.66 & 76.06 & +7.64 \\
  & Java & 44.10 & 54.15 & \best{+22.79} & 42.79 & 52.40 & +22.46 & 57.21 & 59.83 & +4.58 \\
  & Py   & 68.66 & 76.07 & +10.79 & 72.36 & 78.06 & \best{+7.88} & 75.78 & 80.34 & +6.02  \\
\midrule
\multirow{4}{*}{\makecell[l]{\textit{Unified KB}\\\textit{(Cross-Lang.\ Best}\\\textit{Practices + Code)}}}
  & C    & 55.51 & 53.74 & \textbf{$-3.19$}  & 52.86 & 50.66 & \textbf{$-4.16$}  & 60.35 & 53.74 & \textbf{$-10.95$} \\
  & C++  & 74.52 & 79.54 & +6.74             & 71.43 & 77.61 & +8.65             & 70.66 & 58.69 & \textbf{$-16.94$} \\
  & Java & 44.10 & 44.10 & ~0.00             & 42.79 & 44.10 & +3.06             & 57.21 & 54.59 & \textbf{$-4.58$}  \\
  & Py   & 68.66 & 74.64 & +8.71             & 72.36 & 77.49 & +7.09      & 75.78 & 81.77 & +\best{7.90}             \\
\midrule
\multirow{4}{*}{\makecell[l]{\textit{Unified KB}\\\textit{(Cross-Lang.\ Best}\\\textit{Practices Only)}}}
  & C    & 55.51 & 59.91 & +7.93             & 52.86 & 52.86 & ~0.00             & 60.35 & 59.03 & \textbf{$-2.19$} \\
  & C++  & 74.52 & 79.54 & +6.74             & 71.43 & 79.15 & +10.81            & 70.66 & 77.22 & +9.28            \\
  & Java & 44.10 & 46.72 & +5.94             & 42.79 & 44.10 & +3.06             & 57.21 & 55.90 & \textbf{$-2.29$} \\
  & Py   & 68.66 & 73.50 & +7.05             & 72.36 & 74.64 & +3.15             & 75.78 & 77.21 & +1.89            \\
\bottomrule
\multicolumn{11}{l}{}
\end{tabular}
}
\vspace{-2.5em}
\end{table}

\subsubsection{Effect of KB Architecture}
A core design decision in \sol{} is the use of separate, language-specific KBs rather than a single unified KB shared across all languages.
To validate this, Table~\ref{tab:kb-arch} compares four configurations: the default per-language design (best practices and code), a language-specific variant without code snippets (Best Practices Only), a unified KB aggregating best practices and code across all languages (Unified with Code), and a unified KB without code (Unified without Code).

Language-specific KBs clearly outperform unified designs. The \sol{} default delivers the highest SR improvement in 10 of 12 model–language pairs, averaging +22.36\%, versus +8.99\% for Best Practices Only, +4.28\% for Unified without Code, and +0.19\% for Unified with Code.
These comparisons indicate two effects. First, code examples matter: within the language-specific design, removing them drops the average gain from +22.36\% to +8.99\%. Second, unifying across languages hurts most: Unified with Code performs worst, falling below baseline in 5 of 12 configurations. 
We attribute this to cross-language snippets introducing idiom confusion, giving the model conflicting signals for the target's safe practices.

\begin{framed}
\textbf{RQ3 Findings:} Each design axis has a clear default. Dynamic k with a small ceiling matches the best fixed k on SR while cutting token cost at higher ceilings. GPT-5-mini summarization is the only variant whose SR gain justifies its added inference cost. Language-specific KBs are essential: merging knowledge across languages degrades SR below the baseline in several configurations, and within the language-specific design, code examples drive most of the gain over best-practice text alone (+22.36\% vs. +8.98\%).
\end{framed}

\subsection{RQ4: Retrieval Strategy}

\begin{table}[t]
\caption{\sol{} vs.\ naive RAG baselines with $k = 5$.
  Structured fields added to KB for lexical/dense baselines.
  Best SR change $\Delta\%$ per model-language pair in \best{bold}. Ret.M. -- retrieval method}
\label{tab:rq4}
\centering\footnotesize\setlength{\tabcolsep}{4pt}
\resizebox{\columnwidth}{!}{%
\begin{tabular}{l l  r r r  r r r  r r r}
\toprule
\multirow{2}{*}{\textbf{Ret.M.}} &
\multirow{2}{*}{\textbf{Lang}} &
\multicolumn{3}{c}{\textbf{Qwen-Coder}} &
\multicolumn{3}{c}{\textbf{DS-Coder}} &
\multicolumn{3}{c}{\textbf{CodeLlama}} \\
\cmidrule(lr){3-5}\cmidrule(lr){6-8}\cmidrule(lr){9-11}
& & Base & SR & $\Delta\%$ & Base & SR & $\Delta\%$ & Base & SR & $\Delta\%$ \\
\midrule
\multirow{4}{*}{\makecell[l]{\textit{\sol}\\(Mistral)}}
  & C    & 55.51 & 66.52 & \best{+19.83} & 52.86 & 66.96 & +26.67          & 60.35 & 75.33 & \best{+24.82} \\
  & C++  & 74.52 & 88.03 & \best{+18.13} & 71.43 & 85.33 & +19.46          & 70.66 & 88.80 & \best{+25.67} \\
  & Java & 44.10 & 55.45 & \best{+25.74} & 42.79 & 62.88 & +46.95          & 57.21 & 76.86 & +34.35        \\
  & Py   & 68.66 & 76.07 & \best{+10.79} & 72.36 & 76.92 & \best{+6.30}    & 75.78 & 80.91 & +6.77         \\
\midrule
\multirow{4}{*}{\makecell[l]{BM25\\(structured)}}
  & C    & 55.51 & 60.79 & +9.51         & 52.86 & 80.18 & \best{+51.68}   & 60.35 & 65.64 & +8.77         \\
  & C++  & 74.52 & 82.24 & +10.36        & 71.43 & 86.49 & +21.08         & 70.66 & 75.29 & +6.55         \\
  & Java & 44.10 & 52.84 & +19.82        & 42.79 & 66.81 & +56.13          & 57.21 & 75.55 & +32.06        \\
  & Py   & 68.66 & 75.21 & +9.54         & 72.36 & 67.52 & $-6.69$         & 75.78 & 82.34 & \best{+8.66}  \\
\midrule
\multirow{4}{*}{\makecell[l]{TF-IDF\\(structured)}}
  & C    & 55.51 & 59.47 & +7.13         & 52.86 & 78.85 & +49.17          & 60.35 & 65.20 & +8.04         \\
  & C++  & 74.52 & 79.54 & +6.74         & 71.43 & 87.64 & \best{+22.69}   & 70.66 & 74.13 & +4.91         \\
  & Java & 44.10 & 52.84 & +19.82        & 42.79 & 68.56 & \best{+60.22}   & 57.21 & 77.73 & \best{+35.87} \\
  & Py   & 68.66 & 75.50 & +9.96         & 72.36 & 64.67 & $-10.63$        & 75.78 & 78.06 & +3.01         \\
\midrule
\multirow{4}{*}{\makecell[l]{Jina-V3\\(structured)}}
  & C    & 55.51 & 58.15 & +4.76         & 52.86 & 76.65 & +45.01          & 60.35 & 65.64 & +8.77         \\
  & C++  & 74.52 & 80.69 & +8.28         & 71.43 & 85.71 & +19.99          & 70.66 & 77.99 & +10.37        \\
  & Java & 44.10 & 53.28 & +20.82        & 42.79 & 61.14 & +42.88          & 57.21 & 70.31 & +22.90        \\
  & Py   & 68.66 & 76.07 & \best{+10.79} & 72.36 & 55.84 & $-22.83$        & 75.78 & 64.67 & $-14.66$      \\
\bottomrule
\end{tabular}
}
\vspace{-1em}
\end{table}

Table~\ref{tab:rq4} compares \sol{}'s LLM-based retrieval (Mistral) against BM25~\cite{bm25}, TF-IDF~\cite{tf-idf}, and jina-embeddings-v3~\cite{jinav3} as alternative approaches over the same KB at fixed $k\!=\!5$.
\sol{} identifies categories purely by their semantic  $primary\_name$, relying on the LLM to interpret the query's pro\-gram\-ming-domain intent and map it to the related categories. 
Lexical and dense-retrieval methods, however, cannot perform this semantic mapping from a category name alone. 
To give these baselines a fair opportunity, we augment each KB entry with a richer set of structured fields to be matched: \textit{search\_keywords}, 
\textit{nat\-u\-ral\_\-lan\-guage\_\-pat\-terns}, \textit{code\_patterns}, and \textit{best\_practice} descriptions. 
Language scoping is handled via regex-based language detection applied before retrieval, ensuring that candidates are drawn only from entries relevant to the target programming language.
 
Despite this augmentation, \sol{} achieves the best average SR improvement across all 12 model-language pairs.
While BM25 and TF-IDF achieve higher SR on individual configurations (e.g., TF-IDF on Java with DS-Coder), we note that it exhibits severe regressions on other individual configurations (e.g., BM25 on Python with DS-Coder).
Regressions arise because lexical and embedding methods match categories based on surface-form similarity across the augmented fields, but lack the capacity to reason about programming-domain intent in context. The LLM categorizer interprets the full semantic intent of the query and selects categories that are contextually appropriate for the target language.
\begin{framed}
\textbf{RQ4 Findings:} LLM-based retrieval is the most robust strategy, achieving the best mean SR gain (+22.12\% vs. +18.96\% BM25, +18.08\% TF-IDF, +13.09\% Jina-V3) with consistently non-negative improvements across all 12 pairs. Lexical and dense methods reach strong gains on some configurations but suffer severe regressions on others, since they match surface forms rather than reasoning about programming-domain intent.

\end{framed}

\section{Discussion}\label{sec:discussion}
\subsection{Observations}
\subsubsection{\sol{} on commercial models}
Our main experiments use open-weight code models, where security-enhancing retrieval helps most. To check whether the benefit also holds for a stronger commercial model, we performed a small experiment using \sol{} with GPT-5.4-mini as the code generator on CyberSecEval.
We found that \sol{} improves SR (+8.85\% on a baseline SR of $\approx$67\% averaged across the four languages),
showing that even a model with stronger built-in security priors benefits from compact, explicitly injected security guidance. As expected, the absolute gain is smaller than for the small open models.

\subsubsection{On Category-Level Abstraction}
 
Prior work on lost-in-the-middle effects~\cite{lost} and selective retrieval~\cite{REPOFORMER} established that LLM attention degrades with prompt length and irrelevant context.
Instead of asking "\textit{which CWE entries are relevant to this query}", \sol{} asks "\textit{which programming domains does this query touch?}"
We argue that this reframing aligns more naturally with how developers reason about their tasks and with how LLMs attend to injected context.
The security gains over CodeGuarder~\cite{CodeGuarder} are not primarily a function of having richer security knowledge, but of delivering that knowledge in a more useful form.

Also, rather than focusing on optimal KB curation, this work aims to introduce a new direction: demonstrating that high-level, domain-specific security guidance can be more effective than CWE-level guidance.
As a result, the current KB may not cover some programming domains and consequently misses some CWEs. To quantify this, we compared the CWEs appearing in two benchmarks, CyberSecEval and CodeGuard+, against those covered by \sol{}'s KB. We found that 12 of the 33 CWEs in CyberSecEval and 18 of the 42 CWEs in CodeGuard+ are not represented in our KB.
We also do not focus on curating the categories themselves: some categories may be overly broad and could benefit from finer granularity. As a preliminary check, we examined the "Cryptography \& Random Number Generation" category for Python on CyberSecEval. We found that in most cases this category was retrieved and explicitly advised against weak cryptographic and weak PRNG functions, yet the model still failed to avoid them, whereas in Java the model successfully avoided weak cryptographic and PRNG functions. This suggests that the category's entry may need strengthening, or that cryptography-related concepts should be separated into distinct categories in Python. We leave a deeper investigation of this to future work.

To sum up, the KB architecture finding is an empirical demonstration of the context-confusion hypothesis~\cite{lost}, and suggests that knowledge organization is at least as important as knowledge content in security-oriented RACG.

\subsubsection{On Security–Functionality Trade-off}
 
A natural concern with security-oriented prompt augmentation is that constraining LLMs toward secure patterns may come at the cost of functional correctness. 
Our HumanEval results (Table~\ref{tab:functional}) alleviate this concern for \sol{} given improvements in both pass@1 and pass@5.
The likely mechanism is that compact, domain-specific security guidance reinforces correct programming idioms alongside safe ones; the boundaries between "writing correct memory allocation" and "writing safe memory allocation" overlap substantially.
Where tension does emerge, Qwen-Coder on CodeGuard+ (Table~\ref{tab:codeguard}), the model shifts toward defensively structured code that satisfies security, but seemingly deviates from the expected functional output distribution of that benchmark.
We think that this is model-specific rather than an inherent property of the approach.

 \subsection{Threats to Validity}

\subsubsection{Construct validity}
\sol{} is evaluated primarily through Security Rate (SR), which is computed using static analysis tools such as Semgrep, Weggli in CyberSecEval, and CodeQL in CodeGuard+. These tools are useful for detecting syntactic patterns associated with insecure code, but they do not determine whether a vulnerability is reachable, exploitable, or impactful in context, and they only cover the weakness classes encoded in their rule sets. Thus, improvements in SR should be interpreted as evidence of fewer statically detectable weaknesses, not a direct measure of real-world security. This limitation is shared with related work, including CodeGuarder, making comparisons fair, but it also means that stronger validation with dynamic analysis or fuzzing would be valuable in future work. 
\sol's effectiveness depends on the categorizer routing each query to appropriate categories. Our RQ4 results show LLM-based selection is more robust than lexical or dense retrieval, and dynamic k lets queries spanning several concerns receive multiple categories. However, we do not directly measure mis-categorization. Quantifying categorizer error and its downstream effect on SR is left to future work.

\subsubsection{Internal validity}
Another threat concerns whether the observed improvements can be attributed to \sol{} itself rather than to confounding factors. To reduce this risk, we evaluate the same set of models across the same benchmarks and compare against the corresponding baselines under matched settings. However, some results may still be influenced by model-specific behavior, prompt sensitivity, or the interaction between security guidance and the underlying code generator. In particular, the relationship between security and functional correctness is not uniform across all tasks, and small variations in benchmark difficulty or model priors may affect results.

\subsubsection{External validity}

We focus on open-weight code generation models in the 7-16B parameter range, which can be run locally without reliance on proprietary APIs. 
Our evaluation spans C, C++, Python, and Java, and extending the category schema to further languages is a natural next step that requires expert effort to construct language-appropriate vulnerability groupings. Finally, by design CAT-RAG targets high-level categorization and agentic retrieval rather than exhaustive KB curation; building a better KB
with systematic coverage of additional CWEs and alternative grouping strategies is promising future work.

\section{Conclusion}\label{sec:Conclusion}

\sol{} is a security category-aware retrieval-augmented code generation framework for LLM-based code generation. It organizes security knowledge into compact, language-specific programming-domain categories that combine best practices and secure code examples covering multiple related weaknesses. Across CyberSecEval, \sol{} improves security rate across all 12 model-language configurations, with an average gain of 23.57\% in the functional-RACG setting while using about 41.71\% fewer total tokens than the current state-of-art. CAT-RAG also remains resilient to both targeted and intent-agnostic RACG poisoning. These results suggest that knowledge organization is at least as important as knowledge content in security-oriented retrieval-augmented code generation. Future research directions include improving KB, automatically updating the security KB as new CWE entries appear, extending this approach to more languages, and validating \sol{} with dynamic analysis or fuzzing, especially on larger proprietary models.

\section*{Acknowledgments}
The manuscript was edited for grammar and clarity using Grammarly and Claude (Anthropic, Sonnet 4.6, Opus 4.8). Generative AI (Copilot) was used to support code implementation and refactoring and to help build the knowledge bases (KBs). All suggested output was reviewed and refined by the authors, and all research content, including experimental design, results, analysis, and conclusions, is entirely the work of the authors.
\clearpage
\bibliographystyle{IEEEtran}
\bibliography{references}

\end{document}